\documentstyle[twocolumn,prl,aps]{revtex}

\input epsf

\begin{document}

\bibliographystyle{unsrt}

\twocolumn[\hsize\textwidth\columnwidth\hsize\csname
@twocolumnfalse\endcsname

\title{X-ray induced persistent photoconductivity in Si-doped
Al$_{0.35}$Ga$_{0.65}$As}

\author{Yeong-Ah Soh$^1$, G. Aeppli$^1$, Frank M. Zimmermann$^2$,
E. D. Isaacs$^3$, and Anatoly I. Frenkel$^4$}

\bigskip

\address{$^1$NEC Research Institute, Princeton, NJ 08540, $^2$Rutgers
University, Department of Physics and Astronomy and Laboratory for
Surface Modification, Piscataway, NJ 08854, $^3$Bell Laboratories,
Murray Hill, NJ 07974, $^4$Materials Research Laboratory, University
of Illinois at Urbana-Champaign, Urbana, IL 61801}

\date{\today}
\maketitle

\begin{abstract}
We demonstrate that X-ray irradiation can be used to induce an
insulator-metal transition in Si-doped Al$_{0.35}$Ga$_{0.65}$As, a
semiconductor with {\it DX} centers.  The excitation mechanism of the
{\it DX} centers into their shallow donor state was revealed by
studying the photoconductance along with fluorescence.  The
photoconductance as a function of incident X-ray energy exhibits an
edge both at the Ga and As K-edge, implying that core-hole excitation
of Ga and As are efficient primary steps for the excitation of {\it
DX} centers.
A high quantum yield ($\gg 1$) suggests that the excitation is
indirect and nonlocal, due to secondary electrons, holes, and
fluorescence photons.
\end{abstract}
]
\narrowtext

Understanding the microscopic mechanisms of deep-level traps is
important to semiconductor technology, due to their inevitable
presence in many semiconductors.  Deep-level traps can degrade device
performance by acting as recombination centers in optical devices,
such as lasers and light-emitting diodes, or by giving rise to high
resistivity\cite{semicond}.  Examples of deep-level traps are Si
dopants in Al$_{x}$Ga$_{1-x}$As, which act as {\it DX} centers when
$x>0.22$\cite{mooney001}.  {\it DX} centers can be optically excited
to a shallow donor state.  At low temperatures, this state is
metastable because of the resulting structural
rearrangement (Fig.~\ref{llr}{\bf A})\cite{lang635}.
In the ground state, substitutional Si dopants in Ga sites are
displaced from the tetrahedral site resulting in a large
bond-rupturing displacement \cite{chadi873,chadi063}.  In the
excitation process, the Si atom moves to the tetrahedral site, and two
electrons are injected into the conduction band, increasing the
conductivity\cite{chadi873,chadi063}.  At temperatures below $\sim 80$
K the induced photoconductivity is persistent\cite{nelson351}.
Excitation of {\it DX} centers using visible light has been studied
extensively in various semiconductor compounds over the last few
decades\cite{tsubaki663sh,mizuta043,mooney829,baraldi491sh}.  Here, we
demonstrate that irradiation by X-rays can also induce persistent
photoconductivity in Si-doped Al$_{0.35}$Ga$_{0.65}$As.  While
phenomenologically the persistent photoconductivity and its erasure
upon annealing is similar to the X-ray induced metallization observed
in a transition metal oxide\cite{kiryukhin813sh}, the underlying
mechanisms are quite different.

There have been several attempts to determine the local structure of
{\it DX} centers using X-ray absorption fine structure
(XAFS)\cite{sette637sh,mizuta126,ishii672sh,espinosa446sh}.  So far, the
results have been inconclusive.  Conventional XAFS, which monitored
the Se fluorescence of Se-doped AlGaAs, failed to show the large
lattice relaxation\cite{mizuta126}.  Recently, a new method for XAFS,
which measures the energy-dependent X-ray induced photocapacitance,
has been suggested to be site-selective, and therefore, to reveal the
large lattice relaxation\cite{ishii672sh}.  On the other hand, XAFS
studies on CdTe:In claimed that X-rays do not induce photoexcitation
based on the observation that consecutive XAFS scans were
reproducible\cite{espinosa446sh}.  In order to resolve this puzzle, we
studied the mechanism of X-ray excitation of {\it DX} centers by
measuring the energy-dependence of X-ray induced photoconductance
along with fluorescence.  Our key finding is that X-rays are efficient
in inducing photoconductance by exciting the {\it DX} centers.
However, the excitation process is indirect, i.e., core-hole
excitation of host atoms not directly bonded to {\it DX} centers can
lead to the excitation of {\it DX} centers.
Thus, standard X-ray techniques for determining the local structure of
{\it DX} centers are complicated by X-ray induced changes in their
configuration.  Unfortunately, methods relying on monitoring the
ability to change their configuration, e.g. measures sensitive to the
carriers liberated after X-ray photoexcitation, are unsuitable because
of the non-locality of the photoexcitation process.

The sample consisted of a Si-doped Al$_{0.35}$Ga$_{0.65}$As film,
grown using molecular beam epitaxy by Quantum Epitaxial Design.  Over
a semi-insulating GaAs substrate the following layers were grown: a
200 nm thick GaAs buffer layer, a 1.5~$\mu$m thick
Al$_{0.35}$Ga$_{0.65}$As spacer, a 1~$\mu$m thick
Al$_{0.35}$Ga$_{0.65}$As:Si layer, and a 10~nm thick GaAs cap layer.
The doping concentration is estimated to be $5 \times 10^{19}~{\rm
cm}^{-3}$, which was extracted from temperature dependent Hall
measurements\cite{thio802sh}.  

Figure~\ref{TdepG} shows the conductance measured as the sample was
cooled in the dark.
From 300 K down to 115 K, the conductance plotted on a logarithmic
scale as a function of inverse temperature shows a linear
behavior. The steep decrease of the conductance is due to {\it DX}
centers, each of which traps two conduction electrons.
From the slope we deduce a value of 0.23 eV for the binding energy of
the {\it DX} centers\cite{factor2}, which is consistent with
previously reported values\cite{mooney786}.

The X-ray experiments were conducted at beam line X16C at the National
Synchrotron Light Source at Brookhaven National Laboratory.  The X-ray
beam was centered in the middle of the sample and was spread laterally
to $\sim 6$ mm to illuminate the region between the two ohmic
contacts.  The vertical width of the beam was less than 1 mm.  At 300
K, no photoconductivity was detected.  The photoconductivity became
noticeable around 210~K, where the effect was about 1 \%.
Fig.~\ref{timeresp} shows the time-dependent response of the
photoconductance at 160 K to a varying incident photon flux together
with the fluorescence.  The energy of the incident X-ray beam was kept
constant at the As K-edge.  In contrast to the fluorescence, which
responds instantaneously to the incident X-ray flux, the
photoconductance grows more gradually.  In addition, while the
fluorescence response is proportional to the incident flux, the
photoconductance shows saturation.  Both effects follow from the
metastability of the photoexcited {\it DX} centers.  As the beam is
turned on, an excess population of donors in the shallow state builds
up, and so increases the photoconductance.  A finite time is required
to reach the steady-state value for a given incident photon flux, and
to decay thermally to the ground state when the beam is turned
off\cite{lin855sh}.
The saturation effect arises due to the finite number of {\it DX}
centers available for excitation, and becomes more pronounced at lower
temperatures because the excited state becomes longer lived.

Energy scans of the photoconductance along with the fluorescence are
plotted in Fig.~\ref{escan} for both Ga and As K-edges at 180 K.
Here, the photoconductance was plotted after subtracting the dark
current conductance.  As depicted in both K-edge scans, the
photoconductance follows closely the fluorescence scan, with steps in
the photoconductance occurring at the same location as for the
fluorescence.  In addition, the step heights are similar for both K
edges: $8.7 \times 10^{-7}$~S for the Ga K-edge with an incident flux
of $4.15 \times 10^{10}$ photons/s, and $8 \times 10^{-7}$~S for the
As K-edge with $4.23 \times 10^{10}$ photons/s.  This result has
several significant implications.  First, it demonstrates that
core-hole excitation is an efficient primary step for the excitation
of {\it DX} centers.  Second, it suggests that the dominant mechanism
for the X-ray excitation of {\it DX} centers is indirect.  The fact
that the additional photoconductance obtained by opening up a new
core-hole excitation channel is similar for As and Ga shows that
direct proximity of the absorbe to the Si dopant is not required since
the Si dopants are believed to be directly bonded to As, not Ga.

As displayed in Fig.~\ref{ppc}, persistent photoconductivity was
induced at 24 K by illuminating the sample with a defocused Ga K-edge
beam.
After a few minutes' X-ray irradiation, the conductance rose sharply
by more than seven orders of magnitude.  The saturated value of the
conductance was $8.04 \times 10^{-4}$~S, higher than the room
temperature conductance of $2.66 \times 10^{-4}$~S.  At this value,
the sample may be considered metallic.  The photoconductance persisted
even after the beam was turned off and remained constant during the
monitored time, which was 17.5 hrs.  The thermal decay of the
persistent photoconductance, monitored during the warm up process in
the dark (Fig.~\ref{TdepG}), shows an annealing temperature $\sim
100$~K.

The quantum yield {\it Q}, defined as the number of {\it DX} centers
converted into shallow donors per incident photon, can be extracted
from the photoconductance data at 24 K and is given by 
\begin{equation}
{dG/dt \times N_{\rm sat}\over 2G_{\rm sat}I_{\rm o}}.
\end{equation}
Here $dG/dt$ is the initial slope of the conductance, $N_{\rm sat}$
is the total number of photogenerated electrons at saturation, $G_{\rm
sat}$ is the saturated value of the conductance, and $I_{\rm o}$ is
the incident photon flux.  Previous measurements using visible light
have shown that the saturation density of photogenerated carriers in
this sample is $n_{\rm sat}\sim 4 \times
10^{18}$~cm$^{-3}$\cite{thio802sh}.  From the measured values for ${dG
\over dt} =1.33 \times 10^{-5}$~S/s, $N_{\rm sat} \sim 3 \times
10^{13}$, $G_{\rm sat}$, and $I_{\rm o}=2.8 \times 10^{10}$~photons/s,
we obtain $Q \sim 10$.  The fact that for each incident photon a large
number of {\it DX} centers are converted to shallow donors is strong
evidence for our model of indirect excitation of {\it DX} centers.
From this value of {\it Q}, we can judge the extent to which the
excitation process is non-local.  The starting point is to note that
at the Ga K-edge, the penetration depth of the incident photons is
$\sim 15~\mu$m.  Thus, the number of incident photons absorbed in the
1~$\mu$m thick Si-doped layer is 1/15 of the incident photons.
If photon absorption only within the next-nearest neighbor shell of
the Si donor were to lead to excitation, the maximum value for {\it Q}
could be approximated by the donor to host atom density ratio (which
is of the order of $10^{-3}$) times the number of nearest (for As
K-edge) or next-nearest (for Ga K-edge) neighbors, multiplied by the
probability that a photon is absorbed in the active layer ($\sim
1/15$).  This would give $Q \sim 10^{-3}$.  From our result of $Q \gg
1$, it is evident that the excitation is highly non-local and
secondary processes need to be invoked to describe the excitation of
{\it DX} centers.  A comparison of the cross-section for the
excitation of {\it DX} centers ($2 \times 10^{-15}~{\rm cm}^2$) with
the cross-section for core-hole excitation ($3 \times 10^{-20}~{\rm
cm}^2$) further illustrates the efficiency of the former process.

After the initial core-hole excitation, Auger processes or secondary
photon emission occur, which create a cascade of secondary electrons,
holes, and photons (Fig.~\ref{llr}{\bf B}).  Eventually, these
secondary electrons and holes thermalize to the band edges.
The extent of non-locality of the excitation process is governed by
the diffusion length of the secondary photons and holes.  It is
worthwhile to compare the newly discovered phenomenon - excitation
induced by X-rays - to the previous experiments using visible light
since the energy scale of the incident photons differ by more than
three orders of magnitude.  Interestingly, the excitation process is
indirect in both cases.  When using visible light, the most efficient
way to excite {\it DX} centers is by first creating electron-hole
pairs through band-gap excitation.  The electrons and holes diffuse
and thermalize to the band edges.  A hole recombines with a {\it DX}
center, which excites the {\it DX} center into a shallow donor.  As a
consequence an electron is emitted to the conduction band, resulting
in the generation of two electrons in the conduction band for each
band-gap excitation.  Although the X-ray excitation process initially
involves a core-hole excitation, the subsequent processes induced by
the secondarily generated carriers, which eventually thermalize at the
band edges, are the same as in the visible light case.

To summarize, we have investigated the excitation of {\it DX} centers
using X-rays in Si-doped Al$_{0.35}$Ga$_{0.65}$As.  Our key results
are that X-rays are very efficient in inducing photoconductivity and
that the predominant excitation mechanism is indirect and non-local.
Secondary electrons, holes, and photons, which are created following
the original core-hole excitation, are primarily responsible for the
conversion of the {\it DX} centers into their shallow donor state.
Therefore, each absorbed incident photon converts a large number of
{\it DX} centers, resulting in a high quantum yield for the conversion
process.  The extent of non-locality is governed by the diffusion of
the secondary photons and charge carriers.

We gratefully acknowledge valuable discussions with D. J. Chadi,
technical assistance from D. Hines, W. Lehnert, A. Noszko, and T. Thio
for providing us with the sample.  AIF acknowledges support by the DOE
Grant No. DEFG02-96ER45439 through the Materials Research Laboratory
at the University of Illinois at Urbana-Champaign.


\begin{thebibliography}{10}

\bibitem{semicond}
Semiconductors and Semimetals, Vol. 19, edited by R. K. Willardson and A. C.
  Beer, Academic Press (1983).

\bibitem{mooney001}
P. M. Mooney, J. Appl. Phys. {\bf 67}, R1 (1990).

\bibitem{lang635}
D. V. Lang and R. A. Logan, Phys. Rev. Lett. {\bf 39}, 635 (1977).

\bibitem{chadi873}
D. J. Chadi and K. J. Chang, Phys. Rev. Lett. {\bf 61}, 873 (1988).

\bibitem{chadi063}
D. J. Chadi and K. J. Chang, Phys. Rev. B {\bf 39}, 10063 (1989).

\bibitem{nelson351}
R. J. Nelson, Appl. Phys. Lett. {\bf 31}, 351 (1977).

\bibitem{tsubaki663sh}
K. Tsubaki {\it et al.}, Appl. Phys. Lett. {\bf 45}, 1 (1984).

\bibitem{mizuta043}
M. Mizuta and K. Mori, Phys. Rev. B {\bf 37}, 1043 (1988).

\bibitem{mooney829}
P. M. Mooney, M. A. Tischler, and B. D. Parker, Appl. Phys. Lett. {\bf 59},
  1829 (1991).

\bibitem{baraldi491sh}
A. Baraldi {\it et al.}, J. Appl. Phys. {\bf 83}, 491 (1998).

\bibitem{kiryukhin813sh}
V. Kiryukhin {\it et al.}, Nature (London) {\bf 386}, 813 (1997).

\bibitem{sette637sh}
F. Sette {\it et al.}, Phys. Rev. Lett. {\bf 56}, 2637 (1986).

\bibitem{mizuta126}
M. Mizuta and T. Kitano, Appl. Phys. Lett. {\bf 52}, 126 (1988).

\bibitem{ishii672sh}
M. Ishii {\it et al.}, Appl. Phys. Lett. {\bf 74}, 2672 (1999).

\bibitem{espinosa446sh}
F. J. Espinosa {\it et al.}, Phys. Rev. Lett. {\bf 83}, 3446 (1999).

\bibitem{thio802sh}
T. Thio {\it et al.}, Appl. Phys. Lett. {\bf 65}, 1802 (1994).

\bibitem{factor2}
We have included the necessary factor of two reflecting the fact that two
  electrons are ejected per {\it DX} center.

\bibitem{mooney786}
P. M. Mooney, N. S. Caswell, and S. L. Wright, J. Appl. Phys. {\bf 62}, 4786
  (1987).

\bibitem{lin855sh}
J. Y. Lin {\it et al.}, Phys. Rev. B {\bf 42}, 5855 (1990).

\end{thebibliography}

\begin{figure}
\caption{{\bf A} Energy diagram for the large lattice relaxation
model.  The {\it DX} center in the ground state (configuration
coordinate $Q_1$) can be optically excited to the metastable shallow
donor state through an initial photon absorption $E_{opt}$, followed
by lattice relaxation to configuration coordinate $Q_0$.  The barrier
$E_c$ for the {\it DX} center to decay to the ground state, the
binding energy $E_b$, and the barrier to excite the {\it DX} center
$E_e$ are shown in the diagram. {\bf B} (a) Initial core-hole
excitation, followed by either (b) photon emission, or (c) Auger
process, which eventually leads to (d) the collection of secondary
carriers thermalized at the band edges.}\label{llr}
\end{figure}

\begin{figure}
\caption{Conductance as a function of temperature measured in the dark
during cool down process plotted together with the conductance during
warm up after persistent photoconductance was induced at 24
K.}\label{TdepG}
\end{figure}

\begin{figure}
\caption{Effect of X-ray irradiation on the photoconductance as a
function of time.  The different curves correspond to different values
for the photon flux.  The fluorescence responds instantly to the X-ray
irradiation and is proportional to the number of incident photons.
The photoconductance responds with a time delay and is nonlinear in
the photon flux, showing a saturation effect.  
}\label{timeresp}
\end{figure}

\begin{figure}
\caption{Energy-dependent photoconductance measured simultaneously
with the fluorescence yield at 180 K.  Scan {\bf A} and {\bf B}
correspond to the Ga and As K-edge, respectively.  The
photoconductance exhibits an edge at the same location as the
fluorescence for both Ga and As K-edges, implying that core-hole
excitation is an efficient primary step for the excitation of {\it DX}
centers.
}\label{escan}
\end{figure}

\begin{figure}
\caption{Persistent photoconductance induced at 24 K using a defocused
beam with energy at the Ga K-edge.  The conductance rose sharply upon
X-ray irradiation, and saturated at $8.0 \times 10^{-4}$~S.  The
conductance remained persistent after the beam was turned off for more
than 17 h, with no signs of decaying.}\label{ppc}
\end{figure}

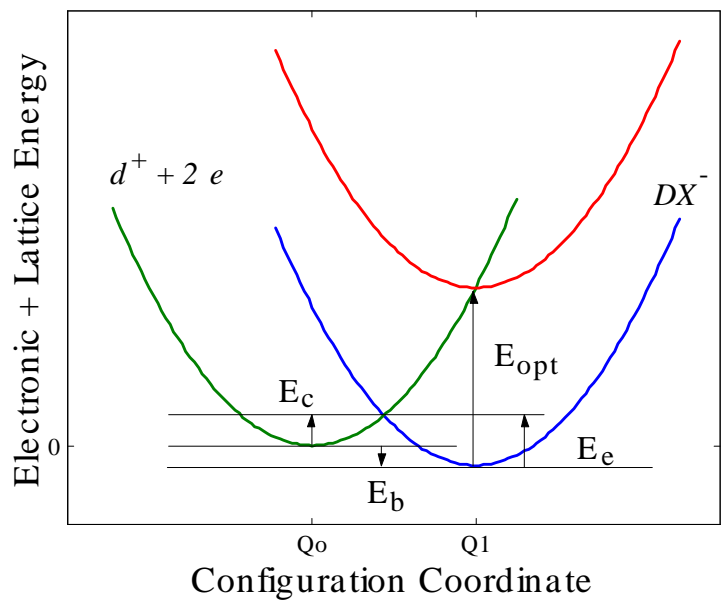
\end{document}